**TRANSPORT AND TELECOMMUNICATION INSTITUTE**

**Helen AFANASYEVA**

# THE ESTIMATION OF TRANSPORT LOGISTIC PROCESSES MODELS ON THE BASE OF INTENSIVE COMPUTER METHODS OF STATISTICS

**SUMMARY OF THE PROMOTION WORK**
to obtain the scientific degree
Doctor of Science in Engineering (Dr.sc.ing.)

Scientific supervisor:
Dr. Habil. Sc. Ing., professor
Alexander ANDRONOV

**Afanasyeva H .**

The Estimation of Transport Logistic Processes Models on the Base of Intensive Computer Methods of Statistics. Summary of the promotion work. – Riga: Transport and Telecommunication Institute, 2006. – 48 p.

**THE PROMOTION WORK PRESENTED TO THE TRANSPORT AND TELECOMMUNICATION INSTITUTE TO OBTAIN THE SCIENTIFIC DEGREE DOCTOR OF SCIENCE IN ENGINEERING (Dr.Sc.Ing.)**

OFFICIAL OPPONENTS

Dr. Habil. Sc. Ing, professor Eugene Kopytov
Transport and telecommunications Institute

Dr. Habil. Sc. Ing, professor Yuri Paramonov
Riga Technical University

Dr. Habil. Sc. Ing, professor Vyacheslav Melas
St. Petersburg State University

The defence of the thesis will be delivered in July 4 2006 at 15 o'clock at the special Promotion Council of Transport and Telecommunication Institute on award of a doctor's degree to the address: 1, Lomonosova street room 4-710, Riga, Latvia, phone (+371) 7100661, fax (+371) 7100660.

CONFIRMATION

I confirm that I developed the promotion work that is presented to the Transport and Telecommunication Institute to obtain the scientific degree Doctor of Science in Engineering. The promotion work has not been presented to any other university or institute to obtain the scientific degree.

May 15, 2006 Helen Afanasyeva

The promotion work is written in English, contains an introduction, 6 chapters, conclusions, 19 figures, 140 formulas and 26 tables, 135 pages in total. Bibliography contains 91 sources.

## ABSTRACT

The promotion work “The Estimation of Transport Logistic Processes Models on the Base of Intensive Computer Methods of Statistics” has been worked out by Helen Afanasyeva to obtain the scientific degree of “Doctor of Science Engineering in Telematics and Logistics”. Scientific supervisor of the work is Dr.habil.sc.ing., professor Alexander Andronov.

The work is devoted to the implementation of the modern statistical methods for the transport and logistic models analysis. The intensive computer method resampling is especially attended. This method is non-parametrical and gives the most efficient estimators of the systems’ characteristics in the case of small initial sample sizes. The investigation was held in three main directions: the forecasting and estimation of logistic models, the estimation of the reliability and efficiency of carries, and the analysis of inventory control problems in logistic systems.

Resampling method usage algorithms and inferences for formulas of the efficiency comparison of traditional and resampling approaches were suggested for each task implementing corresponding mathematical model. The efficiency criteria were: bias, variance or mean squared error of the estimator. The numerical results proving the efficiency of the suggested approach were obtained. The conclusions and recommendations, concerning the conditions in which the suggested approach is the most effective were made. The obtained results are general, because can be used in other subject areas.

# CONTENTS

## 1. Actuality of the Problem

Transport of Latvia is a quickly developing branch recognized as one of the most prior. The given branch demands extensive and expensive scientific researches, with attraction of information technologies, the mathematical apparatus, methods of economics and logistics.

In the given promotion work the greatest attention is paid to the application of modern statistical methods in planning and the organization of logistical processes of transport. The research goes in three basic directions: forecasting and estimation of transport flows; estimation of reliability and efficiency of carries; inventory control of logistics systems.

Solving practical problems with application of stochastic models, we often face such problem, as shortage of statistical data on the basis of which it is necessary to estimate unknown parameters of stochastic models. In such situation use of traditional parametrical methods of estimation is rather complicated. However, intensive computer methods allow to solve this problem. These methods of computing statistics have appeared rather recently. They assume repeated use of the same data in various combinations that provides fuller use of the statistical information. An important point and the integral advantage of this approach is that it is nonparametric. The nonparametric feature of resampling is illustrated here, showing that it is possible to avoid the mistakes peculiar to the traditional methods of hypotheses testing about a kind of distribution of random variables (r.v.). The basic attention in work is paid to the intensive computer method resampling. The comparative scheme of traditional and resampling approaches, with reference to imitating modeling is presented on fig. 1. Here it is illustrated the nonparametric feature of resampling, it is shown, that r.v. are not generated by special generators, but are undertaken directly from initial sources. Resampling allows to estimate stochastic characteristics of complex systems, being based on rather small volumes of statistical data.

As the resampling method is rather new there is an actual problem of research of efficiency of the given approach to various mathematical models and practical problems.

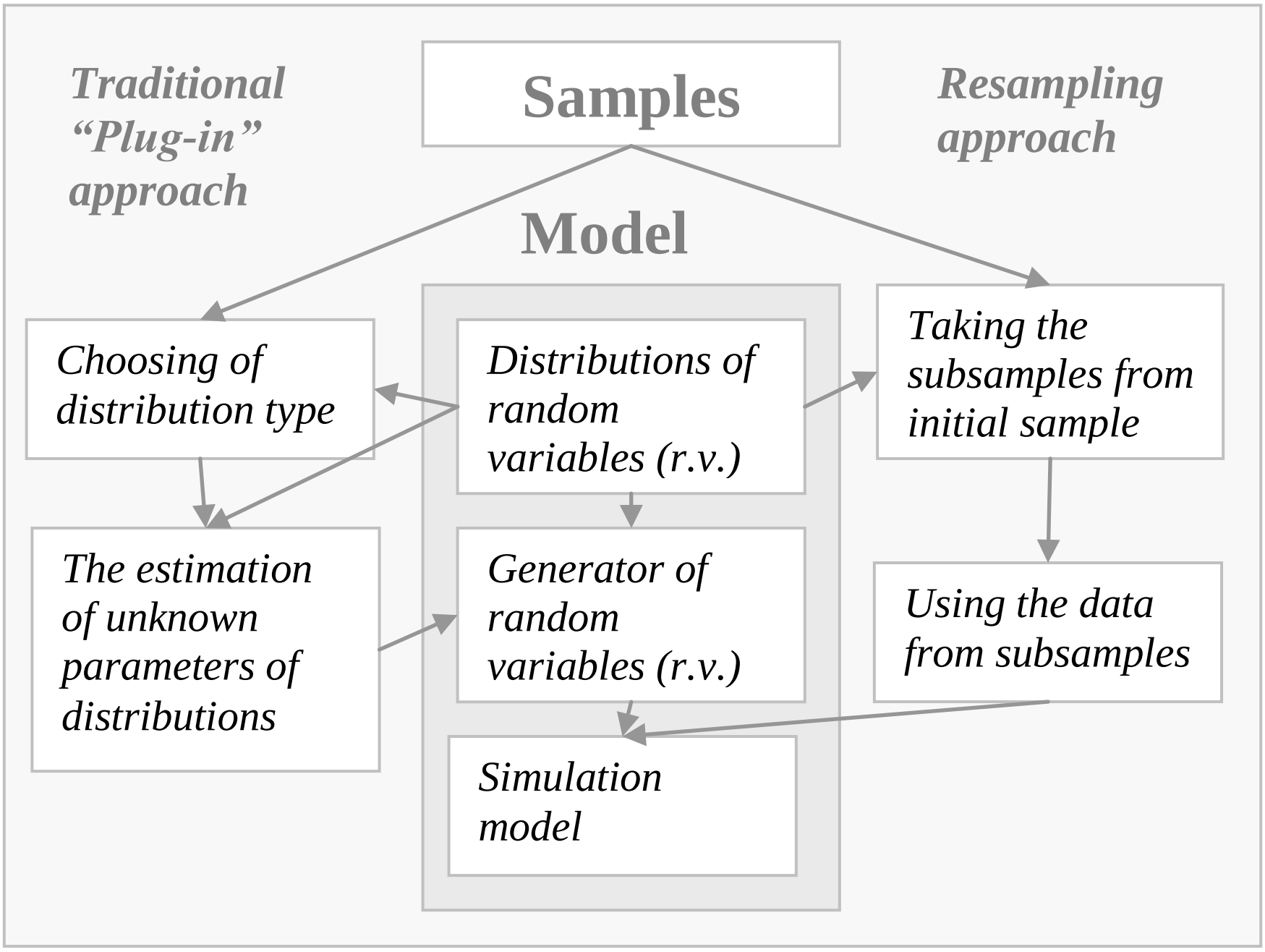

**Fig. 1Comparison scheme of the plug-in and resampling approaches**

## 2. Aim and Tasks of the Research

The main purpose of the research is the development of resampling method usage algorithms for different mathematical models analysis, the estimation of the efficiency of this approach, and also the implementation of it for various practical tasks solving, concerning the estimation of models of logistics processes of transport.

The following issues are supposed to be the main tasks of the work:

- To study the main intensive computer methods of statistics (ICM) and fields of their applications;
- To develop the algorithms of resampling approach application to regression models parameter estimation.
- To investigate the efficiency of resampling-method for regression models construction.
- To consider the possibility of applying the resampling approach for the forecasting and estimation of transport flows on base of regression models.

- To suggest algorithms of using resampling approach for queuing systems parameters estimation.
- To develop algorithms for the method efficiency estimation for statistical tasks of queuing theory.
- To consider the possibility of applying the resampling approach for the estimation of reliability and efficiency of carries on base of the theory of queuing processes.
- To suggest methodic of resampling approach application for the task of comparison of two renewal processes.
- To estimate the efficiency of suggested method for the task of comparison of two renewal processes.
- To apply the resampling approach for inventory control on base of model of the comparison of two renewal processes.
- To develop the software program complex for new approach implementation and its efficiency estimation.
- Using the resampling methodology, to make estimations for different models from the transport systems area and apply algorithms for the estimators' efficiency estimation.

## 3. Readiness of the Theme

Intensive computer methods (ICM) of statistics or computation statistics methods began extremely develop with the appearance of the powerful computers. ICM include the following methods: cross-validation, jackknife method, bootstrap, resampling, which can help in solving wide set of tasks. Jackknife method was suggested by M. Quenouille in 1949. It was the estimator, which was the combination of the estimator's obtained using all data and the estimators, obtained using only part of the same data. B. Efron suggested bootstrap method in 1979, which was the generalization of the jackknife method. In 1976 V. Ivnitsky suggested to use resampling for the estimation of the systems' reliability by simulation. In 1979 C. Wu considered the implementation of the resampling-approach to the estimation of regression model. Resampling and bootstrap began to develop extremely from 1995 under prof. A. Andronov supervision. During this investigation, simple and hierarchical resampling and their implementations in reliability theory, queuing theory and optimization tasks were considered. After that with Yu. Merkuryev and M. Fioshin assistance there were considered the resampling for the sums for partially known distributions, resampling-approach to confidence intervals construction. There were made research works with H. Afanasyeva participation in the branches of resampling approach implementation for the estimation of: order statistics [4], of regression models' parameters [6], [5], renewal theory models, queuing theory models [2], [9]. Some of them were described in the current promotion work. Then H. Afanasyeva considered the resampling-approach implementation for the task of the comparison of renewal processes [2] and its using for the tasks of the inventory control theory [8], which was included as a chapter in the promotion work. The researches in this area are continuing.

## 4. Methodology and Methods of the Research

For a theoretical and methodical basis of promotion's work classical works in the field of mathematical and computational statistics, transport, logistics, modeling and computer sciences are taken. Last tendencies of development of a science in these areas are considered.

In work books in noted areas, thematic materials of periodicals, materials of the international conferences, statistical collections were used.

All conclusions are based on application of classical devices of the theory of stochastic processes, probability theory, and mathematical statistics.

During the research various examples from area of transport logistics for which application of a suggested technique has been illustrated have been analyzed. In examples were used both real and hypothetical data which to the full reflect the specification and efficiency of the applied approach. Bias, variance and mean squared error of estimators has been taken as criteria of efficiency of utilized method and specification of considered practical task. Tendencies of change of efficiency depending on various factors which allow to judge an opportunity of application of a method for the decision of practical problems have been analyzed.

To solve the problems stated in the work, both analytical and experimental methods were applied. By means of the analytical methods analytical expressions for analysis of efficiency of a method in various situations are received. By means of the experimental methods criteria of efficiency of an applied technique for concrete numerical examples which allow to do a conclusion about influence of various factors on the results are counted up.

The results of application of an suggested technique are compared to the results of application of classical approaches. Conclusions about necessary conditions at which resampling-method is more effective are done.

## 5. Scientific Innovation

It is possible to speak about two aspects of scientific novelty of work. Firstly, from the mathematical aspect. This is the application of resampling-method in various statistical tasks and the analysis of efficiency of their application. Secondly, from the applied aspect. Till now the application of the resampling-method to transport systems in the given statement was not analyzed.

In the given promotion work application of modern statistical methods in planning and the organizations of logistical processes of transport in three basic directions are analyzed: forecasting and estimation of transport flows; estimation of reliability and an efficiency of carries; inventory control in logistics systems.

Mathematical basis for the decision of the first problem connected with the forecasting of transport flows was regression model. Within the limits of the given part of research the resampling-approach was applied for estimation of parameters of regression models. Algorithms of application and algorithms of calculation for the analysis of efficiency are new scientific results.

The mathematical model for the second part of research connected with estimation of reliability and efficiency of carries a queuing system became. The application of the resampling-approach to the considered situation have been considered, formulas for the analysis of efficiency are deducted. It is the scientific novelty of the second part of the research.

At last, within the limits of the last part of the research the resampling approach has been applied to the decision of problems of the inventory theory. The mathematical device of the analysis was the processes of inventory control. The problem of an estimation of probability of deficiency in a specific target of inventory control which could mathematically be described as a problem of comparison of two renewal processes was solved. Besides algorithms of application in the given situation of the resampling-approach, expressions for an estimation of efficiency of the given approach, making new scientific results of the last part also have been received.

## 6. Practical value, realization and application of the work

A practical result is applications of the suggested approaches to the decision of practical problems of transport logistics. The first considered problem is a forecasting and estimation of logistical processes as models of some macro-economic activities. Within the limits of this chapter by way of illustration the following problems were solved:

- ❑ Forecasting the passenger flow for air transportation in the countries of the European Union for the certain year, depending on various factors: territory of the country, the population, average wages, and an internal total product per capita.
- ❑ Forecasting the demand for transport service in various regions, depending on an urban saturation, an educational level and a wage level.

The second considered class of problems is connected with estimation of reliability and efficiency of carries. Within the limits of the given chapter the example of the decision of such practical problem is given:

- ❑ There is a flow of damages on an aviation construction. There is a statistics of occurrence of damages before the development of a dangerous situation. It is necessary to estimate probability that in a considered interval of time there will be no refusal.

The third sphere of application of the given technique is an inventory control theory. The approach is illustrated on an example of a following problem:

- ❑ The process of exploitation of some inventory units (fleet of vehicles) is considered. The initial inventory level of these units, and also statistics about intervals between their deliveries and failure of each commodity unit is known. It is necessary to estimate the probability that in the moment of demand for the given product unit, the shortage will not occur, and also to find an optimal inventory level at the set parameters of profit on use, costs on storage and losses from shortage.

The received numerical results allow to judge efficiency of a considered campaign for the decision of the described problems. The analysis of results allows to draw a conclusion on an advisability of application of the resampling-approach in similar problems of transport logistics. The obtained results are general, because can be used in other subject areas.

## 7. Publications

The results of the work were considered in 10 publications in scientific journals and the proceedings of international conferences. The described publications were presented at scientific international conferences in Latvia, Poland, Israel and France.

## 8. Structure of the work

In the first chapter application of statistical methods in planning and the organizations of logistical processes of transport is considered. In the second chapter modern intensive computer methods (ICM) of statistics are considered. In the second chapter resampling-method applications overview for different mathematical models is made. Each of the subsequent chapters considers application of the resampling-approach for the decision of a specific target. These are the problems of forecasting on the basis of regression models of transport processes, estimation of reliability and efficiency of carries, inventory control. Each chapter has problem's description, mathematical model, the analysis of traditional approaches to the decision of the given problem and motivation of the suggested approach. Further the maintenance of the suggested approach with algorithms and expressions for analysis of efficiency is stated. Each chapter comes to an end with the numerical analysis of efficiency of the approach, on the basis of hypothetical data which to the full reflect specificity and features of a considered problem. The example of application from area of the transport, containing statement of a problem, algorithm of application of the suggested approach and numerical results further follows. In the end of each chapter conclusions are drawn concerning expediency of application of the suggested approach in the considered situation. Comparison with results of application of traditional methods is made; recommendations are given, under what conditions application of the suggested approach yields the best results. The work is completed with the conclusion containing the general conclusions of work and the bibliography.

# 9. Description of the main results of the research

## 9.1 Statistical methods in planning and organization of logistic processes of transport

The most significant problems of transport are connected with a qualitative statistical data processing, which allows properly organize, optimize, analyze and forecast different indices of efficiency of these systems. It is necessary to create such mathematical apparatus, which allows to analyze the effectiveness of the work of transportation systems, to organize the process of transport, to forecast the demand of transport, and to work out the optimal variants of the development of the transportation system. Logistics, in its turn, is one of the main fields of scientific and technological development of the transportation sector of European Union. It is as a developing branch of economy and a new scientific trend.

In the current work modern statistical methods implementation for some tasks of the planning and organization of logistic processes of transport are proposed. The actuality of it comes with that, it is not possible to manage complex economic systems without arrangement of trustworthy statistical information about the researching objects.

With the help of the corresponding models analysis the following problems can be solved: to plan more effective routes for a better organization of transport and passenger flows; statistics analysis about road accidents to the find out factors causing the accidents; to simulate programs to maintain the proper condition of the road surface, to forecast the possible road repair, the forecasting of the demand in different modes of transport, the routing of the passengers' flows, roads and airways constriction scheme, reserves planning of fleet of vehicles. The mistakes here lead to the inefficient usage of production facilities and incomplete satisfaction of the demand.

To get the data often is a very complicated task. Described situation features the tasks of reliability of some modes of transport and transport technique. Because of the high reliability of, for example aviation technique, there is a very small statistic of failures.

In this work the problems of forecasting and estimation of transport flows, estimation of reliability and efficiency of carries, inventory control problems in logistic systems are considered. The following mathematical and statistical methods are used: regression models, unknown parameters estimation, statistical hypothesis testing, renewal theory, theory of stochastic processes, queuing theory. Resampling methods of the estimation of the unknown systems parameters are implemented, in comparing with traditional plug-in analogs.

## 9.2 Contemporary intensive computer methods of statistics

The development of classical (traditional) statistical methods was usually bounded by the restrictions to the amount of calculations, which earlier were performed by hands. So the main classical results were connected with the asymptotical approach, for the big sample sizes. For the small sample sizes it was possible to use simple enumeration. The question was opened, how to deal with the intermediate sample size. With the appearance of computers this problem almost disappeared, because the methods of computational statistics (intensive statistical computer methods) began to develop.

The traditional statistical methods and traditional statistical models are often based on many assumptions (linear relations, class of distributions, stationarity etc.). On one hand, such assumptions simplify a model and make possible the application of analytical methods. But on other hand, in practical tasks such assumptions can lead to incomplete and incorrect model. It makes the results obtained by using such methods, inaccurate.

It is difficult to apply traditional methods to the analysis of complex systems, non-stationary systems, to the case when the distribution classes that differ from classical. The analytical models construction for such systems is hard and often impossible. So, in these cases it is better to apply ICM.

Note that the ICM are not precise as classical methods are. But it is true only in the case when assumptions of classical methods take place. However, in the case when assumptions of classical methods are not fulfilled, the ICM are more accurate. So, we can say, that the ICM can solve many problems which could not be properly solved before.

We must remember about a potential risk of using the ICM. We need to understand that a big number of computations does not guarantee that the information is used correctly, and the results will be better. In many cases the classical method can be applied for such task, and it give more precise result than we can obtain after a day of computations using the ICM.

The field of computational statistics at the present time includes a big number of methods. These methods allow to view data from different perspectives, analyze different subsets of data. As the number of different combinations of input data can be very large, a big number of computations can be required.

*Monte Carlo methods*, the first methods, which require a big amount of computations were developed in 60-ies. These methods are randomization of experiments conditions with principally incomplete enumeration, by this experiments themselves can be not only physical experiments, but also computer calculations. Monte Carlo methods can be used for hypothesis testing, for determining the significance level of statistics, calculated on the base of the data sample and for the estimation of the unknown parameters' of the system.

*Randomization methods* are *Cross-validation* and *Jackknife*.Rearranging, subsampling or other manipulations with sample cannot give us additional information. But rearranging the dataset we can obtain information, for example, how unusual is dataset relative to the hypothesis. This is an idea of randomization. *Cross validation* is a procedure in which we divide the sample to ”estimation set” and ”validation set”. The estimation set is used to estimate parameters of interest, but validation set is used to analyze the quality of the model. *Jackknife method* was introduced in 1949 by Maurice Quenouille. The idea was to exclude from consideration the observations sequentially, proceed all remaining information and forecast the result in the excluded point. The set of obtained biases on all points contains information about the total bias, which can be used. Moreover, this data contains also information about variance, which opens new perspectives for this procedure. J. Tukey, which participated actively in perfection of this method, called it the jackknife.

*Bootstrap* initially was proposed by B. Efron in 1977 and then it was generalized by other authors. The core idea of the method is to use available statistical data directly in simulation. Bootstrap supposes repeated processing of different parts of the same data, performing, as it were, there turn «by different sides». Bootstrap is non-parametrical method, which firstly was developed to avoid a bias, produced by the small sample. For this problem solution we need to select the models, where the results weakly depend on the real situation. It is technique of non-parametrical statistics, which leaded to the non-parametrical and robust methods development. It becomes known later, that it can be used for estimation of the sample variance, confidence intervals construction and hypothesis testing.

*Trace-Driven simulation* method is applied for validation of constructed model relatively to real system. Trace-Driven simulation is described in 1998-2001 in works of J. Kleijnen. This author considered also the implementation of this method to the validation of transport models.

## 9.3 Resampling-method in statistics and its development

It was also mentioned that it is very difficult to draw a distinction between the different types of ICM. They are called simply resampling methods in various literary sources.

Resampling is an ICM, which can be applied in statistics and simulation. It is alternative method for systems simulation and it can be effectively used in the case of small samples, in the case when one sample is available for various variables. This approach to system simulation was described earlier in papers by A. Andronov. It was shown there that the Bootstrap method allows to avoid estimate bias and to decrease its variance. The core idea of the resampling-method implementation is, that in the process of the simulation the values of r.v. are not generated, but are extracted from the corresponding sample populations. It can be determined that the sizes of available data are not equal and too small, to perform the presentable simulation. This difficulty can be overcome by resampling-approach implementation. It supposes that the same data for one r.v. can be uses many times in different combinations with other r.v. data.

So, The known function $\phi$ on $m$ independent r.v. $X_1,X_2,...,X_m$: $\phi(X_1,X_2,...,X_m)$ is considered. It is assumed that the distribution function $F_i(.)$ of r.v. $X_i$, is unknown, but the sample population is available for each $X_i$:

$$H_i = \{X_{i1}, X_{i2}, \ldots, X_{in_i}\},\ |H_i| = n_i,\ i=1,..,m. \qquad (1)$$

Our aim is to find the estimation of the expectation:

$$\Theta = E(\phi(X_1, X_2, \ldots, X_m)). \qquad (2)$$

Using this approach the values of arguments $X_i$ of the function $\phi$ are extracted randomly from the corresponding samples $\{H_i\}$. In other words $j(l)=(j_1(l),\ldots,j_m(l))$, $l=1,2,\ldots$, are random samples from $\{H_i\}$. We will say, that $j(l)$ and $X(l)=(X_{1j_2(l)},X_{2j_2(l)},\ldots,X_{mj_m(l)})$ form the $l$-th resample. Then the resampling estimator of the parameter $\Theta$ has the following form:

$$\Theta^* = \frac{1}{r}\sum_{l=1}^{r}\phi(X(l)). \tag{3}$$

Usually the number $r$ is a number of resampling procedure's replications, which is noticeably less, than the number of all possible combinations of initial data.

The resampling-method implementation is shown in the fig. 2. Here we can see, that the vector $j(l)$ consists indices of those elements, of the sample, which were extracted at the $l$-th realization. So, for $l$=1, $j$(1)=(2,1,4,2), it means that we took the second element from the first sample $\{H_1\}$, the first element from the second sample $\{H_2\}$, the forth element from the third sample $\{H_3\}$ and the second element from the forth sample $\{H_4\}$. The elements' selection plan for some resample ir considered to be conservative. We suppose that all the resamples $X(l), l=1, 2, \ldots,$ have the same marginal distribution, that corresponds to the product of initial distributions $\{F_i\}$ and samples $(X(l), X(l'))$ have the same joint distributions for all $l \neq l'$. Obviously, for each conservative plan, the estimator is unbiased: $E(\Theta^*)=\Theta$.

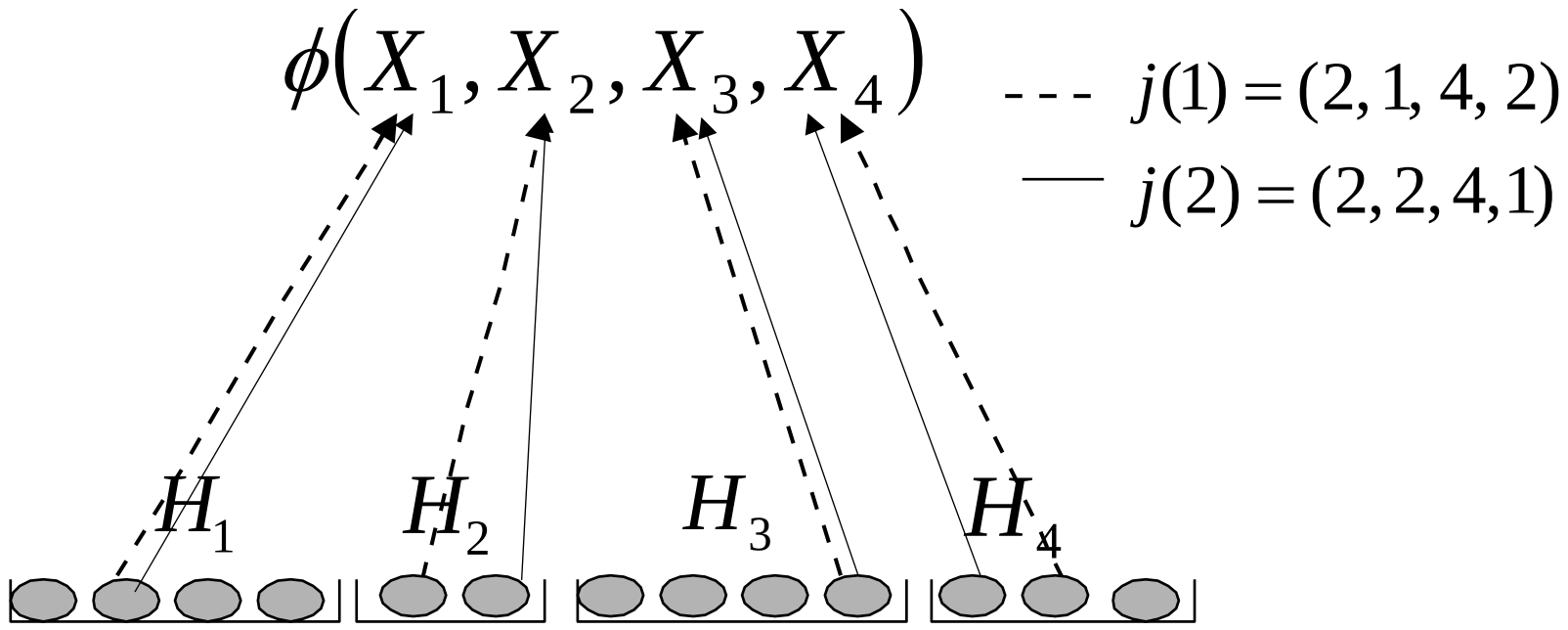

**Fig. 2 Resampling procedure's example**

Note that the conservative plans implicate various versions of resampling approach: simple resampling, controlled (weighted) and hierarchical, stochastic processes resampling and etc., but do not implicate the item-by-item examination.

The different resampling-procedures were described, taking variance of the estimator as efficiency criterion.

We will use the following notations for the moments:

$$\mu_\nu = E(\phi(X(l))^\nu), \quad \nu = 1,2,\ldots, \tag{4}$$

$$\mu_{11} = E(\phi(X(l)),\phi(X(l'))), \quad l \neq l'. \tag{5}$$

Then our variance of interest for estimator can be represented by this way:

$$Var\,\Theta^* = E\Theta^{*2} - \mu^2, \quad E\Theta^{*2} = \frac{1}{r}\left(\mu_2 + (r-1)\mu_{11}\right). \tag{6}$$

Note that $\mu_{11}$ value depends upon the used method of element extraction from the given samples (with replacement, without replacement).

### 9.4 Regression models of transport processes

One of the major initial procedures of formation of the long-term plan is forecasting. Forecasts help to define what processes expediently to develop with advancing rates, allow to prepare alternatives of their development and more or less objectively to compare them.

As the mathematical model for the forecasting time series and regression models can be used.

Let's consider the case of the forecasting on base of regression model. We will describe the resampling-procedures implementation to the regression model's parameters estimation [6].

As it is known, that linear regression model is one of the most popular statistical models. It has the following form:

$$\mathbf{Y} = \mathbf{X}\boldsymbol{\beta} + \mathbf{Z}, \tag{7}$$

where **X** is the $n \times m$ matrix of independent variables, $n$ is the number of observations and $m$ is the number of independent variables, **Y** is the $n \times 1$ vector of dependent variables, **Z** is the $n \times 1$ vector whose components $Z_1$, $Z_2$, ... , $Z_n$ are independent, identically distributed random variables with mean zero and variance $\sigma^2$, so $Cov(\mathbf{Z}) = \sigma^2\mathbf{I}_n$, and $\boldsymbol{\beta}$ is the $m\times 1$ vector of parameters of the regression model to be estimated.

*Classical (traditional) approach*

The classical least squared estimator (LSE) of parameter $\boldsymbol{\beta}$ is well known**:**

$$\widehat{\boldsymbol{\beta}} = \left(\mathbf{X}^T\mathbf{X}\right)^{-1}\mathbf{X}^T\mathbf{Y}. \qquad (8)$$

It is proofed in mathematical statistics, that for the linear regression model the best most effective estimators are obtained by use of the least squared method.

With this estimator (8) in hand we can predict the value of dependent variable $Y$ for selected observation $d$, which is represented by the $1\times m$ vector $\mathbf{x}_d$ of the independent variables values by the following formula:

$$\hat{Y}_d = \mathbf{x}_d \cdot \hat{\boldsymbol{\beta}}. \qquad (9)$$

The estimator (8) can be successfully used if there are no nuisance observations in the sample, otherwise the quality of obtained estimator is not very good, it may be biased. To improve the quality of obtained estimator, we use resampling approach instead of classical one. We investigate suggested method efficiency, taking the bias of the estimators as efficiency criterion.

*The resampling-approach*

Resampling approach works as follows. Let $\mathbf{X} = \left(\mathbf{x}_1^T\ \mathbf{x}_2^T\ \ldots \mathbf{x}_n^T\right)^T$, where $\mathbf{x}_i$ is the $i$-th row of matrix $\mathbf{X}$, that corresponds to the $i$-th observation, $k$ be an integer number, $m \le k < n$, and $N = \{1, 2, ... , n\}$ be the set of integer numbers 1, 2, ... , $n$. We produce a cycle of $r$ steps. In the current (for example the $l$-th) step we form resample by extraction (without replacement) $k$ numbers from $N$: $J(l, 1)$, $J(l, 2)$, ... , $J(l, k)$. So vector $\mathbf{J}(l) = (J(l, 1)\ J(l, 2)\ ...\ J(l, k))$ defines the numbers of the observations (rows of $\mathbf{X}$ and $\mathbf{Y}$), that have been extracted in the $l$-th resample. Then we form the $l$-th resample:

$$\mathbf{X}(\mathbf{J}(l)) = \left(\mathbf{x}_{J(l,1)}^T\ \mathbf{x}_{J(l,2)}^T\ \ldots \mathbf{x}_{J(l,k)}^T\right)^T,$$

$$\mathbf{Y}(\mathbf{J}(l)) = \left(Y_{J(l,1)}\ Y_{J(l,2)}\ \ldots Y_{J(l,k)}\right)^T.$$

It allows us to calculate estimator $\boldsymbol{\beta}^*$ of the model parameter $\boldsymbol{\beta}$ corresponding to (8):

$$\boldsymbol{\beta}^*(\mathbf{J}(l)) = \left(\mathbf{X}(\mathbf{J}(l))^T \mathbf{X}(\mathbf{J}(l))\right)^{-1} \mathbf{X}(\mathbf{J}(l))^T \mathbf{Y}(\mathbf{J}(l)). \tag{10}$$

So, after $r$ replications we have the sequence of estimators:

$$\boldsymbol{\beta}^*(\mathbf{J}(1)), \boldsymbol{\beta}^*(\mathbf{J}(2)), \ldots, \boldsymbol{\beta}^*(\mathbf{J}(r)). \tag{11}$$

Then each component arithmetical mean of this sequence gives resampling-estimator of $\boldsymbol{\beta}$:

$$\boldsymbol{\beta}^*(\mathbf{J}) = \frac{1}{r}\sum_{l=1}^{r} \boldsymbol{\beta}^*(\mathbf{J}(l)). \tag{12}$$

In this chapter the robustness of classical and resampling-estimators in case of disturbed model are investigated, taking bias as obtained estimators' efficiency criterion.

*The definition of the disturbed model*

Let us consider the case when there nuisance observations among all the observations exist. Let us denote "*true*" observations by index $t$ and nuisance, "*false*" ones by index $f$. Without the loss of generality we suppose the true observations to correspond to the first $n$ - $h$ rows of **X**, **Y** and **Z**:

$$\mathbf{Y} = \begin{bmatrix} \mathbf{Y}_t \\ \mathbf{Y}_f \end{bmatrix}, \quad \mathbf{X} = \begin{bmatrix} \mathbf{X}_t \\ \mathbf{X}_f \end{bmatrix}, \quad \mathbf{Z} = \begin{bmatrix} \mathbf{Z}_t \\ \mathbf{Z}_f \end{bmatrix},$$

where $\mathbf{Y}_t = \left(Y_1\, Y_2 \ldots Y_{n-h}\right)^T$, $\mathbf{Y}_f = \left(Y_{n-h+1}\, Y_{n-h+2} \ldots Y_n\right)^T$, $\mathbf{X}_t = \left(\mathbf{x}_1^T\, \mathbf{x}_2^T \ldots \mathbf{x}_{n-h}^T\right)^T$, $\mathbf{X}_f = \left(\mathbf{x}_{n-h+1}^T\, \mathbf{x}_{n-h+2}^T \ldots \mathbf{x}_n^T\right)^T$, $\mathbf{Z}_t = \left(Z_1\, Z_2 \ldots Z_{n-h}\right)^T$, $\mathbf{Z}_f = \left(Z_{n-h+1}\, Z_{n-h+2} \ldots Z_n\right)^T$,

$$\begin{gathered} \mathbf{Y}_t = \mathbf{X}_t\boldsymbol{\beta} + \mathbf{Z}_t, \qquad \mathbf{Y}_f = \mathbf{X}_f\boldsymbol{\beta}_f + \mathbf{Z}_f, \\ E\mathbf{Z}_t = \mathbf{0}, \quad Cov(\mathbf{Z}_t) = \sigma^2\mathbf{I}_{n-h}, \quad E\mathbf{Z}_f = \mathbf{0}, \\ Cov(Z_f) = \sigma^2\mathbf{I}_h, \quad Cov(\mathbf{Z}_t, \mathbf{Z}_f) = \mathbf{0}. \end{gathered} \tag{13}$$

Such model is said to be *disturbed model*.

There were obtained the following estimators of both approaches during the research process, when only one observation if false. The expectation of classical estimator for disturbed model is the following:

$$E\widehat{\boldsymbol{\beta}} = \boldsymbol{\beta} - \frac{1}{1 + \mathbf{x}\left(\mathbf{X}_t^T \mathbf{X}_t\right)^{-1} \mathbf{x}^T} \left(\mathbf{X}_t^T \mathbf{X}_t\right)^{-1} \mathbf{x}^T \mathbf{x}\left(\boldsymbol{\beta} - \boldsymbol{\beta}_f\right). \qquad (14)$$

The expectation of the resampling-estimator for disturbed model.

$$E\boldsymbol{\beta}^* = \boldsymbol{\beta} - \binom{n}{k}^{-1} \times$$

$$\times \left[ \sum_{\mathbf{u} \in \mathbf{L}_t(k-1)} \frac{1}{1 + \mathbf{x}\left(\mathbf{X}^T(\mathbf{u})\mathbf{X}(\mathbf{u})\right)^{-1} \mathbf{x}^T} \cdot \left(\mathbf{X}^T(\mathbf{u})\mathbf{X}(\mathbf{u})\right)^{-1} \right] \mathbf{x}^{\mathbf{T}} \mathbf{x}\left(\boldsymbol{\beta} - \boldsymbol{\beta}_{\mathbf{f}}\right). \qquad (15)$$

We can easy see, that both of them are biased. In the numerical examples the bias will be investigated. The case of various false observations was also described.

*Resampling median estimators of regression model*

This approach is considered in the publication [4].

Let us discuss the procedure of obtaining the resampling median estimator for the regression model. We can not use median approach directly for the sequence (11), because its elements are vectors and it is not possible to order them naturally to find the median of the sequence. That’s why we will employ this approach to the estimation of the predictors. We consider the situation, when we wish to get predictor of dependent variable for some existing or future observation. Using $d$-th resample-estimator (11) for $\mathbf{x}_d$ (vector row) we obtain resample-estimators for predicted value of $Y_d$ by formula (9) for current realization:

$$Y_d^*(\mathbf{J}(l)) = \mathbf{x}_d \boldsymbol{\beta}^*(\mathbf{J}(l)),\ l = 1, \ldots, r, \qquad (16)$$

So for each element of the sequence (11) we obtain the sequence of resample-estimators for predicted value of $Y_d$:

$$Y_d^*(\mathbf{J}(1)), Y_d^*(\mathbf{J}(2)), \ldots, Y_d^*(\mathbf{J}(r)). \qquad (17)$$

Then we order them in increasing order of magnitude:

$$Y_{d(1)}^*, Y_{d(2)}^*, \ldots, Y_{d(s+1)}^*, \ldots Y_{d(r)}^*, \qquad (18)$$

where $r=2s+1$ and the resample-estimator, that caused the middle value $Y^*_{d(s+1)}$ in the ordered sequence, will have the name *resampling median estimator* of regression parameters.

Despite the fact that the average is the unbiased estimator of expectation, but the median estimator is biased, there are some advantages of the last one. Let's consider some aspects, including advantages and disadvantages of using median estimator of $\boldsymbol{\beta}$ instead of the average estimator (12). The first aspect is that unbiasness as criterion has some disadvantages, so unbiased estimators may have "incorrect" sign, big variance, not robust results in case of "noisy" data (our case). The second one is that if we take $E(\Theta^* - \Theta)^2$ as efficiency criterion, the biased estimator may be better. The last one is that, median estimators have the following advantage: they are median unbiased robust estimators, when we deflect from assumptions' statistical hypothesis.

*Numerical efficiency analysis of the both approaches*

To illustrate the efficiency of the proposed approach, let us consider a numerical example of the forecasting some commodity consumption. The data are taken from the book Draper and Smith, with $n$=9 observations and $m$=3 independent variables or factors.

We calculated the Mahalonobis squared distances, that is distance measure between each observation and the center of the observations:

**D**=(2.62 5.281 1.95 0.524 3.686 3.695 1.379 1.205 3.659).

Analyzing it we can draw the conclusion that observations with the numbers 2, 6, 5 and 9 are the most probable candidates to be outliners because of the big distance.

*The comparison of the classical and Resampling estimators*

A number of experiments were performed to compare the quality of classical and resampling- estimators of $\boldsymbol{\beta}$. It is clear that if we have no false observations in our data the classical estimators (8) will be the best in the class of linear unbiased estimators.

Alternatively, if we have outliners in our model, than classical approach estimators would be biased. Suppose that one of our observations be false. Let's calculate the bias of the classical estimators (8) denoted by $Bias\,\hat{\boldsymbol{\beta}} = \boldsymbol{\beta} - E\hat{\boldsymbol{\beta}}$ and the bias of the resampling-estimators (12) denoted by $Bias\,\boldsymbol{\beta}^* = \boldsymbol{\beta} - E\boldsymbol{\beta}^*$.

We will check all possible variants, supposing by turn each observation as false one. Corresponding numerical results are given in table 1. Here the first column denotes the index of false observation, the second column denotes the parameter to be estimated and the third column contains the bias of the classical estimator (8). The next four columns contain the bias $Bias\,\boldsymbol{\beta}^*$ of the resampling-estimator (12). Each column corresponds to resample sizes $k$=3, 5, 6, 7.

The table 1 shows that the resampling-approach gives better results, if the observation 2 and 6 are considered to be “false”. Note, that these observations have the biggest Mahalonobis distance. It allows us to perform the following practical recommendations.

Before the starting of regression analysis we should calculate the Mahalanobis distances of all given observations and determine the ones which have the biggest distance. If the last observations may be false, then we would use resampling-approach for regression model estimation.

*Resampling-median estimators*

Taking the same data we performed another series of experiments. For considering model as the only false observation we take the observation number 2, because it is the most possible candidate in accordance with its distance from the center. Then we obtain the resampling median estimator of predicted value according to formula (18). These results we compare consequently with the results of traditional method taking the bias as the efficiency criterion.

To investigate the resample size $k$ influence on the estimators properties we vary it $m \le k < n$. Note that in our example the number of resamples $r$ for obtaining each resampling estimator is equal to all possible combinations $k$ from $n$. In general case the number of resamples is considerably smaller than all possible combinations.

**Table 1**

**Comparison of the bias of classical and resampling-estimators of regression parameters**

| $i$ | $\boldsymbol{\beta}$ | $Bias\,\hat{\boldsymbol{\beta}}$ | $Bias\,\boldsymbol{\beta}^*$ | | | |
|---|---|---|---|---|---|---|
| | | | $k$=3 | $k$=5 | $k$=6 | $k$=7 |
| 1 | $\beta_1$ | 0,018 | 0,268 | 0,033 | 0,020 | 0,015 |
| | $\beta_2$ | -0,433 | -1,500 | -0,563 | -0,477 | -0,436 |
| | $\beta_3$ | 0,263 | 0,328 | 0,294 | 0,278 | 0,268 |
| 2 | $\beta_1$ | 1,895 | 0,989 | 1,403 | 0,020 | 0,015 |
| | $\beta_2$ | -7,615 | -3,900 | -5,586 | -0,477 | -0,436 |
| | $\beta_3$ | 0,158 | 0,055 | 0,097 | 0,278 | 0,268 |
| 3 | $\beta_1$ | 0,879 | 1,561 | 1,310 | 1,170 | 1,027 |
| | $\beta_2$ | -4,126 | -6,844 | -5,873 | -5,310 | -4,729 |
| | $\beta_3$ | 0,397 | 0,420 | 0,432 | 0,425 | 0,413 |
| 4 | $\beta_1$ | -0,350 | -1,138 | -0,538 | -0,452 | -0,394 |
| | $\beta_2$ | 1,455 | 4,757 | 2,233 | 1,874 | 1,635 |
| | $\beta_3$ | 0,042 | -0,039 | 0,026 | 0,034 | 0,039 |
| 5 | $\beta_1$ | 0,171 | 0,243 | 0,194 | 0,178 | 0,171 |
| | $\beta_2$ | -1,190 | -1,771 | -1,363 | -1,258 | -1,204 |
| | $\beta_3$ | 0,307 | 0,414 | 0,339 | 0,322 | 0,312 |
| 6 | $\beta_1$ | -0,541 | 0,063 | -0,122 | -0,266 | -0,266 |
| | $\beta_2$ | 3,527 | 0,678 | 1,646 | 2,311 | 2,311 |
| | $\beta_3$ | -0,576 | -0,364 | -0,465 | -0,510 | -0,510 |
| 7 | $\beta_1$ | -0,978 | -1,773 | -1,395 | -1,240 | -1,099 |
| | $\beta_2$ | 4,386 | 7,636 | 6,125 | 5,477 | 4,886 |
| | $\beta_3$ | -0,172 | -0,278 | -0,235 | -0,210 | -0,188 |
| 8 | $\beta_1$ | 0,056 | 0,981 | 0,421 | 0,259 | 0,139 |
| | $\beta_2$ | 0,046 | -3,401 | -1,296 | -0,706 | -0,262 |
| | $\beta_3$ | -0,053 | -0,058 | -0,065 | -0,057 | -0,054 |
| 9 | $\beta_1$ | -1,067 | -1,111 | -1,224 | -1,168 | -1,116 |
| | $\beta_2$ | 5,286 | 5,681 | 6,013 | 5,751 | 5,510 |
| | $\beta_3$ | -0,410 | -0,521 | -0,467 | -0,444 | -0,426 |

We obtained the resampling median estimators of the predicted values of dependent variable for all existing observations and compare the results with classical estimators, and clean estimators, calculated without false observation. The last means that we removed the false observation from our model and use classical LSE estimator (8) for our purposes. The results are presented in the table 2.

The first column contains the names of parameters of interest the predictors for all observations. The second column contains the bias - the difference between the classical estimator and the clean estimator for all observations. In the next columns we can see the results of implementing resampling median estimators approach for different resample sizes *k*. It contains the bias – the difference between the resampling median estimator and clean estimators for all observations. Analyzing the obtained results we can conclude that for all observations there are the resampling estimators with the smaller value of bias, than the classical one, choosing the right resample size.

Especially good results were obtained for the resample size equal to 6. For example, for the 7-th observation classical approach gives the bias 1.519, but the resampling median estimators are in the interval of (0.416-1.572) depending on the resample size.

**Table 2**

**The bias of resampling median estimators for the example**

| Par. title | The bias of estimators of predictors | | | | | | |
|---|---|---|---|---|---|---|---|
| | LSE | Resampling- median | | | | | |
| | | $k$=3 | $k$=4 | $k$=5 | $k$=6 | $k$=7 | $k$=8 |
| $Y_1$ | 0.115 | 0.617 | 0.43 | 0.332 | 0.139 | **0.034** | 0.333 |
| $Y_2$ | 5.793 | 7.57 | 7.441 | 5.802 | **4.85** | **5.037** | **5.631** |
| $Y_3$ | 1.957 | 5.131 | 3.165 | 2.301 | **1.702** | **1.682** | 1.967 |
| $Y_4$ | 0.497 | 2.571 | **0.108** | **0.243** | 0.545 | 0.739 | 0.772 |
| $Y_5$ | 0.129 | 1.012 | 0.636 | 0.282 | **0.024** | **0.042** | **0.066** |
| $Y_6$ | 1.481 | 4.385 | 3.53 | **1.242** | **0.523** | **0.517** | **1.246** |
| $Y_7$ | 1.519 | **0.416** | **0.681** | **0.959** | **1.204** | 1.572 | 1.935 |
| $Y_8$ | 1.199 | 1.557 | 1.523 | **0.111** | **0.073** | **0.539** | **0.749** |
| $Y_9$ | 0.499 | 2.857 | 0.956 | **0.261** | **0.493** | 0.543 | 0.778 |

*The example of regression model for the forecasting the passengers transportations by air.*

Let's consider the example of the regression model for the forecasting the passengers transportation by air transport. The forecasting is based on the following factors for each country during a given year:

- $x_1$ - the square (km$^2$);
- $x_2$ - population (persons);
- $x_3$- gross domestic product (EUR);
- $x_4$ - the average month salary of the workers (EUR).

The corresponding regression model will have the following way:

$$y=\beta_0+\beta_1 x_1+\beta_2 x_2+\beta_3 x_3+\beta_4 x_4+\varepsilon,$$

where

$y$ - is a number of transported by given country during given year,

$\varepsilon$ - is a random component, that has normal distribution with mean equals 0 and constant variance.

The initial data for the forecasting were obtained from the official Web-site of the Statistical data of European Union. Those data are presented in the work. The first part of the data was used for model construction; the second part was used directly for the forecasting and for model validation.

For the correctness of the classical model the following criteria were used: Fisher's criterion for the testing of the hypothetic about the nonsignificance of the regression (with significance level $\alpha$=5%); Student's criterion for the testing of the hypothesis about the nonsignificance of the *i*-th accompanying variables ($\alpha$=5%); the multiple determinacy coefficient $R^2$.

We decided further to make some transformation with the variables, using some combination of factors. It is shown in models 2-3. The individual populations' mobility (the proportion of transported passengers and the common population) of country $y/x_2$ is considered.

All models' coefficients were estimated using resampling and traditional approaches. The models themselves, their coefficients and the multiple determination coefficients are presented in the table 3.

Resampling-approach almost always (for all models) gave comparable or even better estimators of the forecasting values, taking as efficiency criterion unbiasedness of the estimators of forecasts. The percentage ratios of those resamples are shown in the table 4.

**Table 3**

**Different models of the forecasting**

| Nr. | Regression model | $R^2$ |
|---|---|---|
| 1 | $y = \beta_0 + \beta_1 x_1 + \beta_2 x_2 + \beta_3 x_3 + \beta_4 x_4 + \varepsilon$,<br>Coefficients, in importance increasing order: $\beta_3$, $\beta_2$, $\beta_{4.}$ | 0.85 |
| 2 | $y / x_2 = \beta_0 + \beta_1 x_4 + \beta_2 (x_1 / x_2) + \beta_3 (x_3 / x_2) +$<br>$+ \beta_4 x_3 /(x_2 \cdot \sqrt{x_1}) + \beta_5 x_3 /(x_2 \cdot x_1) + \varepsilon$<br>Coefficients, in importance increasing order: $\beta_1$, $\beta_3$, $\beta_5$, $\beta_4$, $\beta_2$, $\beta_{0.}$ | 0.90 |
| 3 | (Excluding Luxembourg)<br>$y / x_2 = \beta_0 + \beta_1 x_3 + \beta_2 x_4 + \beta_3 (x_1 / x_2) + \beta_4 (x_3 / x_2) +$<br>$\beta_5 (x_4 / x_2) + \beta_6 \sqrt{x_1} + \beta x_3 /(x_2 \cdot \sqrt{x_1}) +$<br>$+ \beta_5 x_3 /(x_2 \cdot x_1) + \varepsilon$<br>Coefficients, in importance increasing order: $\beta_7$, $\beta_3$ ,$\beta_{8,}$ $\beta_2$ | 0.92 |

**Table 4**

**Comparison of forecasting results for traditional and resampling approaches**

| Nr. | Model data (resampling/traditional) | Validation data (resampling/traditional) |
|---|---|---|
| 1 | 24/14<br>(63/37)% | 17/7<br>(71/29)% |
| 2 | 21/17<br>(55/45)% | 15/9<br>(62.5/37.5)% |
| 3 | 25/10<br>(71/29)% | 5/14<br>(26/74)% |

*Inferences*

The resampling-approach to the linear regression model estimation has been considered in the case where the false observations, belonging to another regression equation, exist among all the observations. The bias of parameters estimators has been calculated for classical and resampling-approach. The conditions under which the resampling-approach gives smaller value of bias have been discussed.The application of the proposed approach to the regression model with nuisance observations leads to obtaining good results. The analysis of the numerical results shows that the considered resampling-median approach gives the better estimators than the classical methods, if we take the bias of expectation $E(Y_d)$ as the estimators' efficiency criterion.

So this chapter illustrates the application of resampling-approach to the forecasting of volumes of the passengers transportation by air.

## 9.5 On some problem of the estimation of the reliability and efficiency of carries

*Problem description*

Reliability is the probability of the fact that the device performs its functions according to the made demand, during determined time interval. As it is known this characteristic is of great importance especially in transport technique. The problems, connected with reliability of transport facilities can be very essential. Problems connected with the reliability of the delivery of goods in logistic chains may cause big financial lacks. That all requires the attracting of mathematical apparatus of the estimation and forecasting to make the right decisions in business.

The limited number of initial statistical data is very typical for such problem, for example the data about coming of order of some device. It is difficult to implement traditional methods of parameters estimation in this situation. It requires the attraction of the new intensive computer methods of statistics to overcome those problems. This investigation results were described in the following works [2], [9].

*Mathematical model and the methods of its analysis*

Let's consider the following mathematical model of the queuing theory, which is typical for reliability theory. The model supposes two types of failures – *initial* and *terminal failures*. The initial failures (or the damages) appear according to homogeneous Poisson process with the rate $\lambda$. Each initial failure degenerates into a terminal failure after a random time *B*. So if an initial failure appears at time $\tau_i$ then the terminal failure appears at the instant $B_i + \tau_i$. The terminal failure and the corresponding initial failure are eliminated instantly. We assume that $\{B_i\}$ are mutually independent identical distributed random variables, independent on $\{\tau_i\}$. Let *F*(*x*) be the distribution function of *B*. We take interest in the number of initial failures *X*(*t*) at time *t* (which did not degenerate to the terminal failures) and the number of terminal failures *Y*(*t*) that have been occurred till time *t*. Let $\Lambda(t) = EX(t)$ and $\tilde{\Lambda}(t) = EY(t)$ be the corresponding expectations, $P_i(t) = P\{X(t) = i\}$, $R_i(t) = P\{Y(t) = i\}$ be the corresponding probability distributions, *i* = 0, 1, … .

It is well known that *X*(*t*) and *Y*(*t*) are mutually independent random variables, by that:

$$\Lambda(t) = \lambda \int_0^t (1 - F(x))dx, \quad \tilde{\Lambda}(t) = \lambda \int_0^t F(x)dx, \tag{19}$$

$$P_i(t) = \frac{1}{i!}(\Lambda(t))^i \exp(-\Lambda(t)), \; i = 0, 1, \ldots \tag{20}$$

The probability $R_i(t)$ is calculated analogously by formula (20) where $\Lambda(t)$ is replaced by $\tilde{\Lambda}(t)$.

In fact the rate λ and the distribution function *F*(*x*) are unknown. We need to estimate $\Lambda(t)$, $\tilde{\Lambda}(t)$, $R_i(t)$ and $P_i(t)$ using the sample of the intervals between initial failures appearances $A_1, A_2, \ldots, A_k$ and the sample $B_1, B_2, \ldots, B_l$.

We consider two methods of the estimation: traditional “plug-in” and resampling. This chapter contains the investigation properties of the expectations and the variances of considered estimators. We take as the efficiency criteria the bias and variance of both approaches. It is shown that the resampling-estimators have some advantages for the small sample sizes *k* and *l*.

*Traditional(plug-in) approach contents*

Plug-in estimators uses estimators $\hat{\lambda}$ and $\hat{F}(t)$ instead of the unknown λ and *F*(*t*). Here $\hat{F}(t)$ is the empirical distribution function of B that has been calculated on the base of $B_1, B_2, \ldots, B_l$, $\hat{\lambda}$ is the point estimator of the rate λ:

$$\hat{\lambda} = \left( \frac{1}{k} \sum_{i=1}^{k} A_i \right)^{-1}. \tag{21}$$

In this case we have the following estimators of $\Lambda(t)$ and $P_i(t)$:

$$\hat{\Lambda}(t) = \hat{\lambda} \int_0^t \left(1 - \hat{F}(x)\right) dx, \tag{22}$$

$$\hat{P}_i\ (t) = \frac{1}{i!} \hat{\Lambda}(t)^i \exp\left(-\hat{\Lambda}(t)\right),\ i = 0,\ 1,\ \ldots\ . \tag{23}$$

The estimators of $\tilde{\Lambda}(t)$ and $R_i(t)$ are calculated analogously. In order to investigate the statistical properties of these estimators, it is necessary to know the distributions of random variables $\hat{\lambda}$ and $\int_0^t (1 - \hat{F}(u)) du$. The formulas for the expectation and variance for plug-in estimators (22), (23) were also obtained. Mean squared error of the estimator $\hat{\Lambda}(t)$ is the following: $MSE\hat{\Lambda}(t) = Var\hat{\Lambda}(t) + \left(E\hat{\Lambda}(t) - \Lambda(t)\right)^2$.

*Resampling approach contents*

The resampling-approach supposes the ordinary simulation procedure with the only difference, that it does not use a generator of random numbers, but it extracts necessary random variable directly from the given sample populations $\{A_1, A_2, \ldots, A_k\}$ and $\{B_1, B_2, \ldots, B_l\}$ at random. Let $k \le l$.

We produce *r* independent realizations of simulated process. On the *q*-th realization we extract elements from $\{A_1, A_2, \ldots, A_k\}$ without replacement, form sequence of the time intervals between the initial failures appearances $A(q) = \{A_{i_1(q)}, A_{i_2(q)}, \ldots, A_{i_{N_t}(q)}\}$ and calculate $\tau_i(q)=\sum_{u=1}^{i} A_{i_u(q)}$ , $i$ = 1,2, …, $N_t(q)$, where $N_t(q)$ is a number of initial failures till the time $t$ for the $q$-th realization:

$$N_t(q)=\begin{cases}\max\{j:\tau_j(q)\le t\} & if \quad \tau_k(q)\ge t,\\ k & otherwise.\end{cases} \tag{24}$$

Analogously we produce the sequence $B(q)=\{B_{j_1(q)},B_{j_2(q)},\ldots,B_{j_{N_t}(q)}\}$ of intervals of initial failure degeneration to terminal failure. Then we calculate the sequence $\{\tau_1(q)+B_{j_1(q)},\tau_2(q)+B_{j_2(q)},\ldots,\tau_{N_t}(q)+B_{j_{N_t}(q)}\}$ of terminal failures times for the $q$-th realization.

Let $\zeta_j(t)$ be the indicator function of the event: "The $j$-th initial failure occurred, but didn't degenerated into terminal failure till the time moment $t$":

$$\zeta_{j,q}(t)=\begin{cases}1 & if \quad \tau_j(q)\le t<\tau_j(q)+B_{j_j(q)},\\ 0 & otherwise.\end{cases} \tag{25}$$

Then the number of initial failures $\mathrm{X}_q(t)$ which didn't degenerate into terminal failures till time $t$ for the $q$-th realization is calculated as:

$$\mathrm{X}_q(t)=\sum_{j=1}^{N_t(q)}\zeta_{j,q}(t)=\sum_{j=1}^{k}\zeta_{j,q}(t). \tag{26}$$

The resampling-estimator of $\Lambda(t)$ is the following:

$$\Lambda^*(t)=\frac{1}{r}\sum_{q=1}^{r}X_q(t). \tag{27}$$

Analogously the number of initial failures which had been degenerated into terminal failures till time $t$ for the $q$-th realization $Y_q(t)$ is calculated by formula (26) replacing function $\zeta_{j,q}(t)$. The resampling-estimator $\widetilde{\Lambda}^*(t)$ can be found by formula (27) replacing $X_q(t)$ by $Y_q(t)$.

Now we need to calculate the resampling-estimators of the probabilities $P_i(t)$ and $R_i(t)$. Let $\Phi_i(X(t))$ be the indicator function of the event $\{X(t)=i\}$. The resampling-estimators of probabilities $P_i(t)$ is the following:

$$P_i^*(t) = \frac{1}{r}\sum_{q=1}^{r}\Phi_i(\mathrm{X}_q(t)). \tag{28}$$

The resampling-estimator $R_i^*(t)$ can be found by formula (28) replacing $\Phi_i(X(t))$. Let us calculate the expectations of resampling-estimators. Obviously, $EP_i^*(t) = E\Phi_i(t)$.

Therefore, the expectation $E\Lambda^*(t)$ of the resampling-estimator $\Lambda^*(t)$ is calculated as follows:

$$E\Lambda^*(t) = q_1\sum_{j=1}^{k} j\frac{(\lambda t)^j}{j!}\cdot\exp(-\lambda t) + q_1 k\cdot\sum_{j=k+1}^{\infty}\frac{(\lambda t)^j}{j!}\cdot\exp(-\lambda t) \;. \tag{29}$$

We also can find the expectation $EP_i^*(t)$ of the estimator $P_i^*(t)$:

$$\begin{aligned} EP_i^*(t) = {} & \sum_{j=i}^{l}\frac{(\lambda t)^j}{j!}\cdot\exp(-\lambda t)\cdot\binom{j}{i}\cdot q_1{}^i\cdot(1-q_1)^{j-i} + \\ & + \binom{l}{i}\cdot q_1{}^i\cdot(1-q_1)^{l-i}\cdot\sum_{j=l+1}^{\infty}\frac{(\lambda t)^j}{j!}\cdot\exp(-\lambda t),\;\; i = 0,1,2,..., \end{aligned} \tag{30}$$

where $\binom{j}{i}$ is the binominal coefficient and $q_1$ – is the probability, that at moment $t$ the considered initial failure still will be initial.

The expectations $ER_i^*(t)$ and $E\tilde{\Lambda}^*(t)$ of the estimators $R_i^*(t)$ and $\tilde{\Lambda}^*(t)$ can be calculated analogously. The expressions for variance of the resampling-estimator (28) were also obtained. Mean squared error can be found as follows: $MSE\Lambda^*(t) = Var\Lambda^*(t) + \left(E\Lambda^*(t) - \Lambda(t)\right)^2$.

*The numerical efficiency analysis of the suggested approach*

*Example: Triangular distribution, as the time of initial failure degenerates into a terminal failure.*

Let's consider the Poisson flow of the initial failures with parameter λ=0.5 and the triangle distribution of degeneration times to terminal failures with the parameters $a$=2:

**Table 5**

**The expectations $E\hat{P}_i(5)$ of the plug-in estimators and $EP_i^*(5)$ of resampling-estimators**

| *i* | *l=3* | | *l=4* | | *l=5* | | *l=8* | | Real |
|---|---|---|---|---|---|---|---|---|---|
| | Plug. | Res. | Plug. | Res. | Plug. | Res. | Plug. | Res. | |
| 0 | .346 | .379 | .348 | .370 | .350 | .368 | .352 | .368 | .368 |
| 1 | .291 | .392 | .307 | .374 | .317 | .369 | .334 | .368 | .368 |
| 2 | .169 | .189 | .176 | .189 | .180 | .186 | .186 | .184 | .184 |
| 3 | .088 | .040 | .087 | .058 | .085 | .062 | .080 | .061 | .061 |
| 4 | .045 | | .041 | .009 | .037 | .014 | .031 | .015 | .015 |
| 5 | .024 | | .019 | | .016 | .002 | .011 | .003 | .003 |
| 6 | .013 | | .010 | | .007 | | .004 | | .001 |
| 7 | .008 | | .005 | | .003 | | .001 | | |
| 8 | .005 | | .003 | | .002 | | .001 | | |
| 9 | .003 | | .001 | | .001 | | | | |
| 10 | .002 | | .001 | | | | | | |

Table 5 presents the expectations of plug-in estimators $E\hat{P}_i(t)$ (Plug.) and of resampling estimators $EP_i^*(t)$ (Res.) of the probability of the interest for the fixed time moment $t$=5, with different resamples sizes $l$, comparing with real probabilities values. We consider the case when the both resamples sizes $l$ and $k$ are equal. We can see, that with increasing of value $l$ the bias of both approaches estimators from real probability becomes less and less and for resampling-estimators disappears totally.

We can also easily compare those estimators using corresponding charts in fig. 3-4. In the fig. 4 we see, that there is the same curve for the real probability and the expectation of the resampling-estimator. Various examples are examined with different distribution types and parameters values, but the results tendency was almost the same.

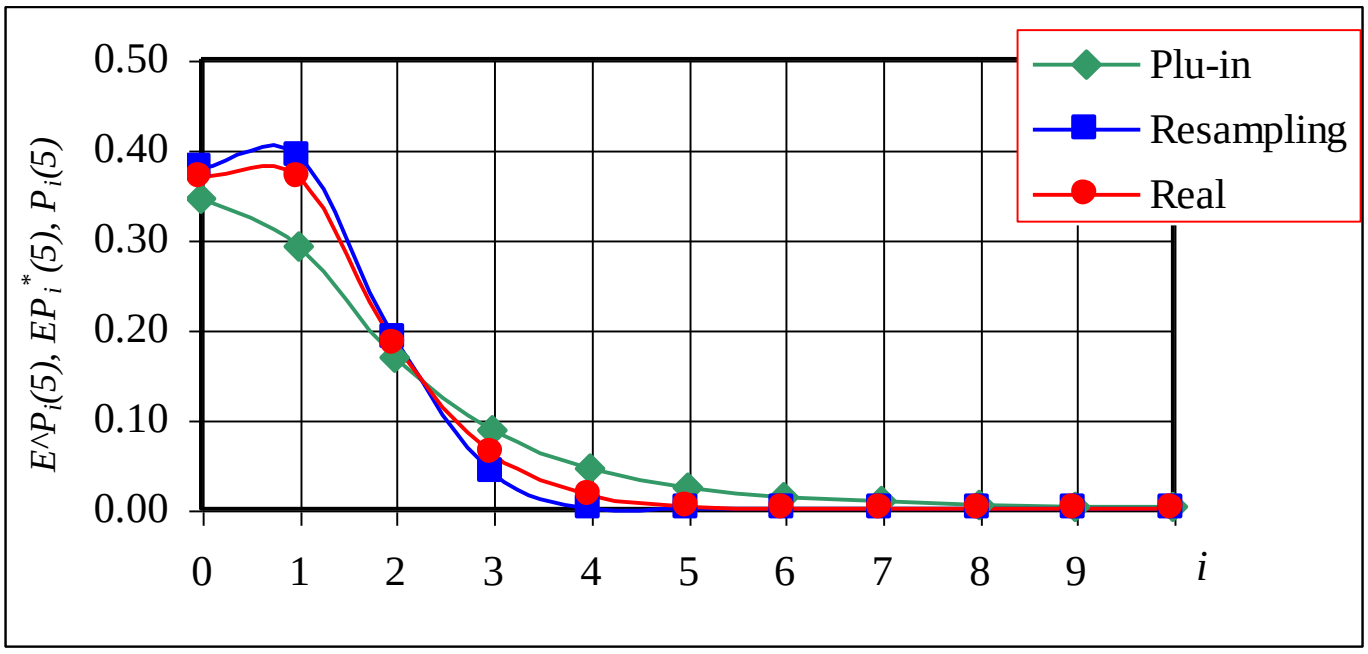

**Fig. 3 The expectations $E\hat{P}_i(5)$ of the plug-in estimator and $EP_i^*(5)$ of the resampling-estimator, *l*=3, *t*=5.**

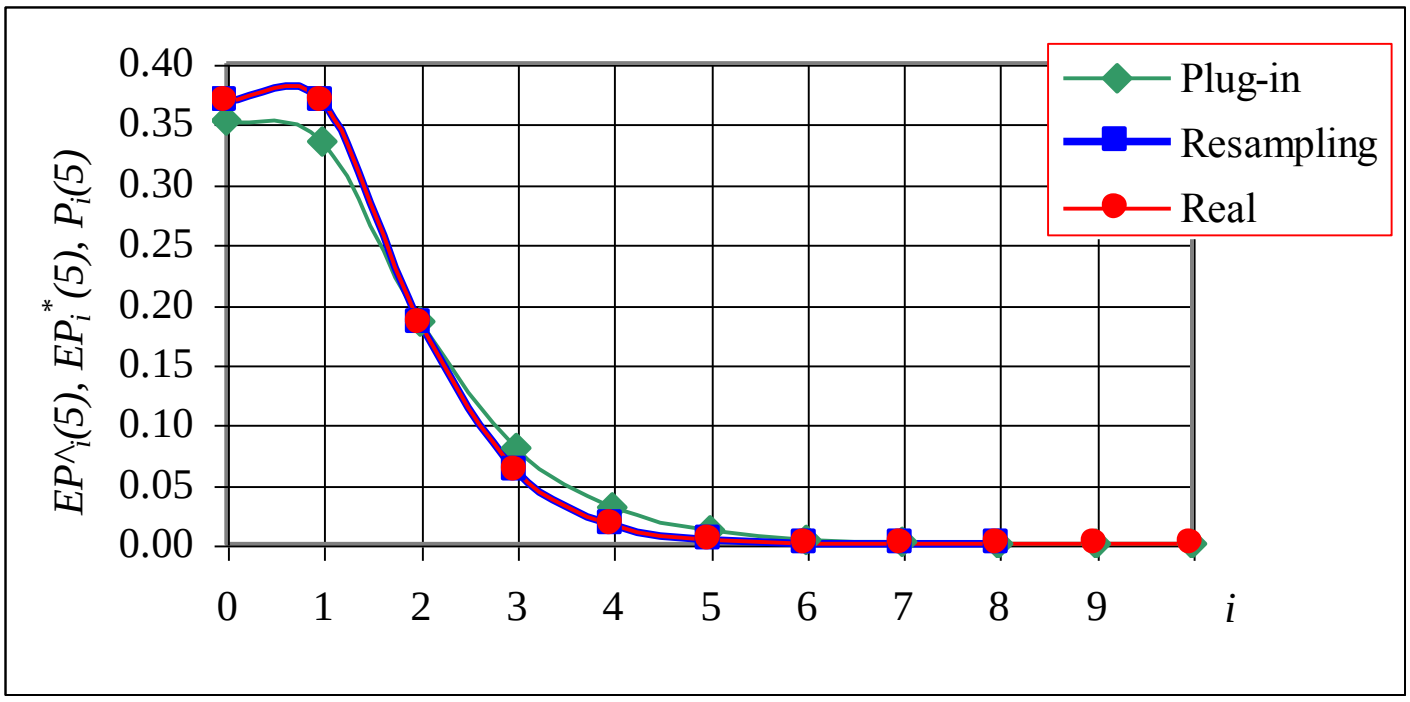

**Fig. 4 The expectations $E\hat{P}_i(5)$ of the plug-in estimator and $EP_i^*(5)$ of the resampling-estimator, *l*=8, *t*=5.**

В таблице 6 находятся математические ожидания $E\hat{\Lambda}(t)$, $E\Lambda^*(t)$, дисперсии $Var\hat{\Lambda}(5)$, $Var\Lambda^*(5)$ и среднеквадратические ошибки $MSE\ \hat{\Lambda}(5)$, $MSE\ \Lambda^*(5)$.

In all cases resampling-estimators gave better expectations, than traditional plug-in ones. It was especially noticeable, when the sample size increased value 3. It was because of the fact that resampling-approach could not give reliable results since, we could not obtain the probabilities of the queue length more than *k* (sample size). But such situation could be in real life. But resampling-approach in many cases gave less bias, than plug-in one. When the sample size was more that 8 the bias of resampling approach disappeared completely.

**Table 6**

**Expectation, variance and mean squared error of the plug-in and resampling estimators of $\Lambda(t)$**

| $i$ | $l$=3 | $l$=4 | $l$=5 | $l$=6 | $l$=7 | $l$=8 |
|---|---|---|---|---|---|---|
| $E\hat{\Lambda}(5)$ | 1.41 | 1.32 | 1.25 | 1.21 | 1.19 | 1.16 |
| $E\Lambda^{*}(5)$ | 0.89 | 0.96 | 0.99 | 0.997 | 0.99 | 0.99 |
| $Var\hat{\Lambda}(5)$ | 1.52 | 0.79 | 0.51 | 0.38 | 0.30 | 0.24 |
| $Var\Lambda^{*}(5)$ | 0.58 | 0.55 | 0.49 | 0.43 | 0.36 | 0.31 |
| $MSE\ \hat{\Lambda}(5)$ | 1.69 | 0.89 | 0.57 | 0.42 | 0.34 | 0.27 |
| $MSE\ \Lambda^{*}(5)$ | 0.59 | 0.55 | 0.49 | 0.43 | 0.36 | 0.31 |

*Inferences*

The proposed resampling-approach is a good alternative to traditional one for considered reliability problem. It is especially remarkable with increasing the size of given samples. Then the rate of convergence to real probability of resampling estimators is much more, than of traditional ones. The only disadvantage of suggested approach is that we cannot get the required value of $EP_i^{*}(t)$, if $i>l$, that's why is those cases is better to use traditional estimators. The best way here is to combine those approaches, to use for $i<l$ resampling estimators, but for $i>l$ traditional ones. We can use such kind of combinations with special normalization to get the sum of probabilities equal to 1.

It is important to emphasize that, when the distribution is estimated correctly. In real situation often the sample size could be too small, that it is impossible to choose the distribution type properly by traditional approach. We can build wrong model, obtain absolutely wrong results and make incorrect decision. Especially in such situations resampling-approach is recommended. If resampling can be competitive even in better situation, when the distributions were estimated correctly, than in described example the advantages of suggested approach are obvious.

## 9.6 Inventory control in logistic systems

*Problem description and mathematical formulation*

If we consider the inventory as the object of the control in logistic systems, then the key questions are connected with determining of optimal level of the inventory, taking into account the risks of inventory holding and shortage appearances.

Suppose we have two simple independent renewal processes $\{X_i, i=1,2,...\}$ and $\{Y_i, i=1,2,...\}$, where $\{X_i\}$ and $\{Y_i\}$ are the sequences of nonnegative independent random variables, each sequence with its own common distribution. Let $D_m = \sum_{i=1}^{m} X_i$ and $S_m = \sum_{i=1}^{m} Y_i$ be the times of the $m$-th renewal for corresponding processes. The distribution functions of sequences $\{X_i\}$ and $\{Y_i\}$ are unknown, but corresponding initial samples' of sizes $n_X$ and $n_Y$ are available. Our purpose is the estimation of the probability $P\{D_m > S_k\}$, where $n_X \geq 2m$ and $n_Y \geq 2k$.

This problem has a lot of applications, for example, in inventory theory it occurs in the following situation. Suppose that the initial inventory level equals to $K$, where $K$ is a known integer. Inventory level is increasing according to the supply and decreasing according to the demand. It is also assumed, that if the demand exceeds the supply then the shortage occurs. Our purpose is to estimate the shortage absence probability for the $m$-th unit's demand.

*Formal problem description*

The described example can be considered in terms of renewal processes in the following way. Let the demand corresponds to the first renewal process $\{X_i, i=1,2,...\}$ and the time of the $m$-th renewal be the time of the $m$-th request of inventory unit. Let the supply corresponds to the second renewal process $\{Y_i, i=1,2,...\}$ and the time of the $m$-th renewal be the time of the $m$-th supply of inventory unit. Then the probability of interest, of the shortage absence, is the probability, that the $m$-th demand comes later, that the $m-K$-th supply $D_m > S_{m-K}$. It is also assumed, that the initial inventory level $K$ is known. We wish to investigate some properties of the different estimators of the shortage absence probability.

Now we describe our problem more formally. The distribution functions $F_X^1(x)$ and $F_Y^1(x)$, of sequences $\{X_i\}$ and $\{Y_i\}$ are unknown, but corresponding samples $H_X = \{X_1, X_2, \ldots, X_{n_X}\}$ and $H_Y = \{Y_1, Y_2, \ldots, Y_{n_Y}\}$ are available, where $|H_X| = n_X$ and $|H_Y| = n_Y$.

We are interested in the time of the *m*-th and $m-K$-th renewals: $D_m = \sum_{i=1}^{m} X_i$, $S_{m-K} = \sum_{i=1}^{m-K} Y_i$. Our task is to estimate the shortage absence probability $P\{D_m > S_{m-K}\}$ that the *m*-th renewal of the demand process $\{X_i\}$ comes later, than the $m-K$-th renewal of the supply process $\{Y_i\}$.

Let's consider the indicator function $\Psi(\mathbf{x},\mathbf{y})$, where $\mathbf{x} = (x_1, x_2, \ldots, x_{m_X})$ and $\mathbf{y} = (y_1, y_2, \ldots, y_{m_Y})$ are vectors of real numbers:

$$\Psi(\mathbf{x},\mathbf{y}) = \begin{cases} 1 & if \sum_{i=1}^{m_X} x_i > \sum_{i=1}^{m_Y} y_i \\ 0 & otherwise. \end{cases} \qquad (31)$$

Suppose we have two vectors of r.v. $\mathbf{X} = (X_1, X_2, \ldots X_{m_X})$ and $\mathbf{Y} = (Y_1, Y_2, \ldots Y_{m_Y})$, $m_X = m$, $m_Y = m - K$. Our purpose is to estimate the shortage absence probability $\Theta = E(\Psi(\mathbf{X},\mathbf{Y}))$. We will estimate $\Theta$ using two different approaches: classical and resampling. Classical, parametrical approach is widely known. So we consider the alternative nonparametric resampling-approach implementation.

*Classical approach*

Classical approach to the estimation of the probability of interest is a parametrical one. It supposes the point estimation of the parameters of the distribution, if we know the distribution type of the initial samples $H_i$, $i=\{X, Y\}$

*Example: Exponential distribution*

Let's consider an example, when r.v. *X* and *Y* have exponential distribution with parameters $\lambda$ and $\nu$ correspondingly. As it is known, that the sum of exponentially distributed r.v. has Erlang distribution. The probability of interest is $\Theta = P\{D_{m_X} > S_{m_Y}\}$.

The classical approach supposes using the point estimators instead of the values of $\lambda$ and $v$:

$$\hat{\Theta} = \sum_{i=0}^{m_X-1} \frac{\hat{v}^{m_Y}}{(\hat{\lambda}+\hat{v})^{m_Y+i}} \frac{\hat{\lambda}^i}{i!} \prod_{p=0}^{i-1} (m_Y + p), \tag{32}$$

where $\prod_{p=0}^{-1} = 1$, $\hat{\lambda} = n_X / D_{n_X}$ and $\hat{v} = n_Y / S_{n_X}$ .

Now we are able to calculate the expectation and the variance of $\hat{\Theta}$ .

*Example: Normal distribution*

Now let's consider the case, where $X$ and $Y$ have normal distribution, correspondingly $N(\mu_X,\sigma_X)$ and $N(\mu_Y,\sigma_Y)$. The real probability of the shortage absence in this case can be calculated as follows: $\Theta = P\{D_{m_X} > S_{m_Y}\} = P\{D_{m_X} - S_{m_Y} > 0\}$ .

If we try to estimate this probability, the classical approach supposes the estimation of the parameters $\boldsymbol{\mu} = (\mu_X, \mu_Y)$, $\boldsymbol{\sigma} = (\sigma_X, \sigma_Y)$ using available sample populations. We have the estimator:

$$\hat{\Theta} = \hat{\Theta}(\hat{\boldsymbol{\mu}}, \hat{\boldsymbol{\sigma}}) = 1 - \Phi\left( \frac{0 - (m_X \hat{\mu}_X - m_Y \hat{\mu}_Y)}{\sqrt{m_X \hat{\sigma}_X^2 + m_Y \hat{\sigma}_y^2}} \right). \tag{33}$$

Now we are able to calculate the expectation $E\,\hat{\Theta}$, the variance $Var\,\hat{\Theta}$ and the mean squared error $MSE\,\hat{\Theta} = Var\,\hat{\Theta} + (\Theta - E\,\hat{\Theta})^2$ of $\hat{\Theta}$ .

*Resampling - approach*

This method in contrast to traditional approach does not suppose the estimation of the distribution parameters or the construction of the empirical distribution functions to find characteristics of interest. Alternatively we use primary data in different combinations and this fact makes possible to obtain unbiased estimators and decrease their variance. Resampling-approach supposes the following steps. We choose randomly $m_X$ elements from the sample $H_X$ and $m_Y$ elements from the sample $H_Y$. The elements are taken without replacement, we remind that $n_X \geq 2 \cdot m_X$, $n_Y \geq 2 \cdot m_Y$. Then we calculate the corresponding value of the function $\Psi(\mathbf{x},\mathbf{y})$ using the formula (31). After that we return chosen elements into the corresponding samples.

We repeat this procedure during $r$ realizations. Let $j_d^i(l)$, $d$=1,..,$m_i$ be the indices of elements from the sample $H_i$, $i \in \{X,Y\}$, that are chosen at the $l$-th realization. Then for the $l$-th realization we obtain the following vectors:

$$\mathbf{X}(l) = (X_{j_1^X(l)}, X_{j_2^X(l)}, \ldots, X_{j_{m_X}^X(l)}),\ \mathbf{Y}(l) = (Y_{j_1^Y(l)}, Y_{j_2^Y(l)}, \ldots, Y_{j_{m_Y}^Y(l)}).$$

The resampling-estimator $\Theta^*$, which is the arithmetical mean by $r$ realizations, is calculated by the formula (3), replacing $\phi(X(l))$ with $\Psi(\mathbf{X}(l),\mathbf{Y}(l))$.

Obviously this estimator is unbiased. We are interested in the variance of this estimator. The moments and variance are calculated using the formulas (4-6) substituting function $\phi(X(l))$ with $\Psi(\mathbf{X}(l),\mathbf{Y}(l))$

In order to estimate the variance of the estimator, we have firstly to find the expression of the mixed moment $\mu_{11}$ from the formula (5). To calculate the moment $\mu_{11}$ the notation of α-pairs can be used.

Let us denote $W_i(l)$, $l$=1,..., $r$, $i \in \{X,Y\}$, a subset of the sample $H_i$, which was used for producing the values of vectors **X**($l$) and **Y**($l$) correspondingly, $W_i(l) \subset H_i$. Let us denote $M_i$={0,1,...,$m_i$}, $M=M_X \times M_Y$. Let $\boldsymbol{\alpha}=(\alpha_X, \alpha_Y)$ be an element of $M$, $\boldsymbol{\alpha} \in M$. We say that $W_i(l)$ and $W_i(l')$ produce the $\boldsymbol{\alpha}$-pair, if and only if $W_i(l)$ and $W_i(l')$ have $\alpha_i$ common elements: $|W_i(l) \cap W_i(l)| = \alpha_i$.

Let $A_{ll'}(\boldsymbol{\alpha})$ denote the event "subsamples (**X**($l$),**Y**($l$)) and (**X**($l'$),**Y**($l'$)) produce $\boldsymbol{\alpha}$-pair", but $P_{ll'}(\boldsymbol{\alpha})$ be the probability of this event: $P_{ll'}(\boldsymbol{\alpha})=P\{A_{ll'}(\boldsymbol{\alpha})\}$. Because of the fact realizations $l$=1,..,$r$ are statistically equivalent, we can omit the lower indices $ll'$ and write $P(\boldsymbol{\alpha})$.

Let

$$\mu_{11}(\boldsymbol{\alpha}) = E\big(\Psi(\mathbf{X}(l),\mathbf{Y}(l))\Psi(\mathbf{X}(l'),\mathbf{Y}(l')) \,|\, A_{ll'}(\boldsymbol{\alpha})\big), \qquad (34)$$

then

$$\mu_{11} = \sum_{\boldsymbol{\alpha} \in M} P(\boldsymbol{\alpha})\mu_{11}(\boldsymbol{\alpha}). \qquad (35)$$

Therefore we need to calculate P{$\boldsymbol{\alpha}$} and $\mu_{11}(\boldsymbol{\alpha})$ for all $\boldsymbol{\alpha} \in M$. The probability P{$\boldsymbol{\alpha}$} can be calculated using hypergeometrical distribution**.** The formula for $\mu_{11}(\boldsymbol{\alpha})$, $\forall \boldsymbol{\alpha} \in M$ was also obtained.

*Numerical affectivity analysis of both approaches*
*Example: Normal distribution*

Consider the case when r.v. *X* and *Y* have normal distribution with parameters $\mu_X = \mu_Y = 2$, $\sigma_X = \sigma_Y = 1$. Let our sample sizes be equal $n = n_X = n_Y$. We consider the mentioned probability at the moment of the *m*-th unit's demand depending on different initial inventory levels *K*=0..3. All calculations have performed for *r* = 1000 realizations.

We intend to compare the variance of estimators of resampling-approach with the mean squared error of classical approach. It is so because of resampling-approach estimators are unbiased, but classical ones on the contrary have bias.

In the table 7 we can see the resampling-estimators' variance $Var\,\Theta^*$ comparing with classical approach estimators' variance $Var\,\hat{\Theta}$, bias $Bias\,\hat{\Theta}$, and mean squared error $MSE\,\hat{\Theta}$. The table shows how changes the results depending on different sample sizes *n*, unit's number *m* and initial inventory level *K*.

Analyzing table's results we can draw the conclusion that the variance and corresponding mean squared error of both approaches decreases with the increasing of sample sizes *n*, *m*, and initial inventory level *K*. The variance of resampling-estimators is almost always near the traditional one. However resampling-estimators are unbiased. Taking as the criterion the mean squared error resampling gives even better results for big values of *K*.

The case where r.v. *X* and *Y* have exponential distribution with parameters λ=0.3, ν=0.7 were also considered. Two approaches were also implemented to estimate the same probability, as in previous example. The criteria of efficiency were the same. Resampling approach showed also good results, sometimes even better, than traditional one.

**Table 7**

**Experimental results for Classical $\hat{\Theta}$ and Resampling $\Theta^*$ estimators**

| | | $K$=0 | $K$=1 | $K$=2 | $K$=3 |
|---|---|---|---|---|---|
| $n$=10 $m$=5 | $Var\,\hat{\Theta}$ | .061 | .043 | .015 | .002 |
| | $Bias\,\hat{\Theta}$ | 0 | .028 | .029 | .013 |
| | $MSE\,\hat{\Theta}$ | .061 | .044 | .015 | .002 |
| | $Var\,\Theta^*$ | .087 | .055 | .014 | .001 |
| $n$=10 $m$=4 | $Var\,\hat{\Theta}$ | .053 | .032 | .006 | --- |
| | $Bias\,\hat{\Theta}$ | 0 | .021 | .018 | --- |
| | $MSE\,\hat{\Theta}$ | .053 | .033 | .007 | --- |
| | $Var\,\Theta^*$ | .069 | .039 | .005 | --- |
| $n$=12 $m$=6 | $Var\,\hat{\Theta}$ | .06 | .045 | .019 | .004 |
| | $Bias\,\hat{\Theta}$ | 0 | .028 | .033 | .019 |
| | $MSE\,\hat{\Theta}$ | .06 | .046 | .02 | .004 |
| | $Var\,\Theta^*$ | .085 | .058 | .02 | .002 |

*Example from the logistic area*

Let's consider the example of the determining of the average income value of the received commodity units with the following initial data:

- $K$- initial inventory level;
- $f(k)$ – the cast of initial inventory level ordering. (for example, $f(k)=b_0+b_1\cdot k$);
- $c_d$ – the income from the fulfillment of the one demand;
- $c_s$ – the penalty of the delay of fulfillment of the demand.

The average income from the $m$ units demand fulfillment:

$$\Pi(m,K) = m \cdot c_d - \left[ f(K) + c_s \cdot \sum_{i=1}^{m} P_i(K) \right], \tag{36}$$

where $P_i(K)$ is the probability that with the initial inventory level $K$ in the moment of the $i$-th unit demand the shortage occurs.

Because of the resampling approach gives the unbiased estimators of the function value, the expectation of this estimator coincides with the real parameter's value. There was found the probability of the shortage absence at the moment of the *i*-th units demand 1-$P_i(K)$ was obtained in the previous part of this chapter. It is clear that the probability of the opposite event can be obtain as follows: $P_i(K)$.

The average damage in the process of *m* units demand fulfillment:

$$\Xi(m,K) = f(K) + c_s \cdot \sum_{i=1}^{m} P_i(K). \qquad (37)$$

Therefore: $\Pi(m,K) = m \cdot c_d - \Xi(m,K)$. The task comes to the maximization of the incomes or the minimization of the damages:

$$\max_K \Pi(m,K) = m \cdot c_d - \Xi(m,K) \text{ or } \min_K \Xi(m,K) = f(K) + c_s \cdot \sum_{i=1}^{m} P_i(K).$$

Let's analyze this function's changing depending from *K* with the fixed values of other parameters $c_d=2$, $c_s=5$, $b_0=0$, $b_1=0.2$, the normal distribution of the demand *N*(2,1), the normal distribution of the supply *N*(2.5,0.2).

It is clear, that in the real situation the distribution of the demand and supply are unknown. But we can estimate them on base on available sample population. We can also estimate the probability of the shortage and the average income from the fulfilled units demand.

Traditional "plug-in" and resampling methods are described in the present chapter. It was noticed, that if the sample sizes are small it is advisable to use resampling-approach. It is so because this approach gives unbiased estimators of parameters with the less mean squared error (variance) value. Let's demonstrate the bias of the traditional approach estimating the average income value. In table 8 the corresponding results ($\Pi$- real income value and $\hat{\Pi}$ - classical estimator) changing the inventory volumes *K* and time (the moment of the *m*-th unit demand).

We can see from the obtained results, that searching the optimal value of the inventory initial level we can mistaken, because of the bias of the classical estimator. In the fig 5 is shown, that in the moment of the 5-th unit demand the optimal initial inventory level is 3 units, but using classical estimator of shortage probability we can obtain the optimal initial inventory level in the point of 4 units. As it is known, resampling-approach gives unbiased estimators of the parameters, so it allows avoiding the noticed disadvantages of the classical approach.

**Table 8**

**Bias of the classical estimator of the average income from the real income value**

| | | $K$=0 | $K$=1 | $K$=2 | $K$=3 | $K$=4 |
|---|---|---|---|---|---|---|
| $m$=1 | $\Pi(1, K)$ | -2.816 | 1.8 | 1.6 | 1.4 | 1.2 |
| | $\hat{\Pi}(1, K)$ | -2.872 | 1.8 | 1.6 | 1.4 | 1.2 |
| $m$=3 | $\Pi(3, K)$ | -9.723 | 2.781 | 5.444 | 5.4 | 5.2 |
| | $\hat{\Pi}(3, K)$ | -11.197 | 2.709 | 5.375 | 5.4 | 5.2 |
| $m$=5 | $\Pi(5, K)$ | -17.624 | 0.462 | 8.02 | 9.285 | 9.197 |
| | $\hat{\Pi}(5, K)$ | -19.9 | 0.34 | 7.604 | 9.144 | 9.18 |
| $m$=7 | $\Pi(7, K)$ | -26.139 | -4.109 | 8.532 | 12.572 | 13.129 |
| | $\hat{\Pi}(7, K)$ | -28.814 | -3.995 | 7.653 | 11.954 | 12.945 |

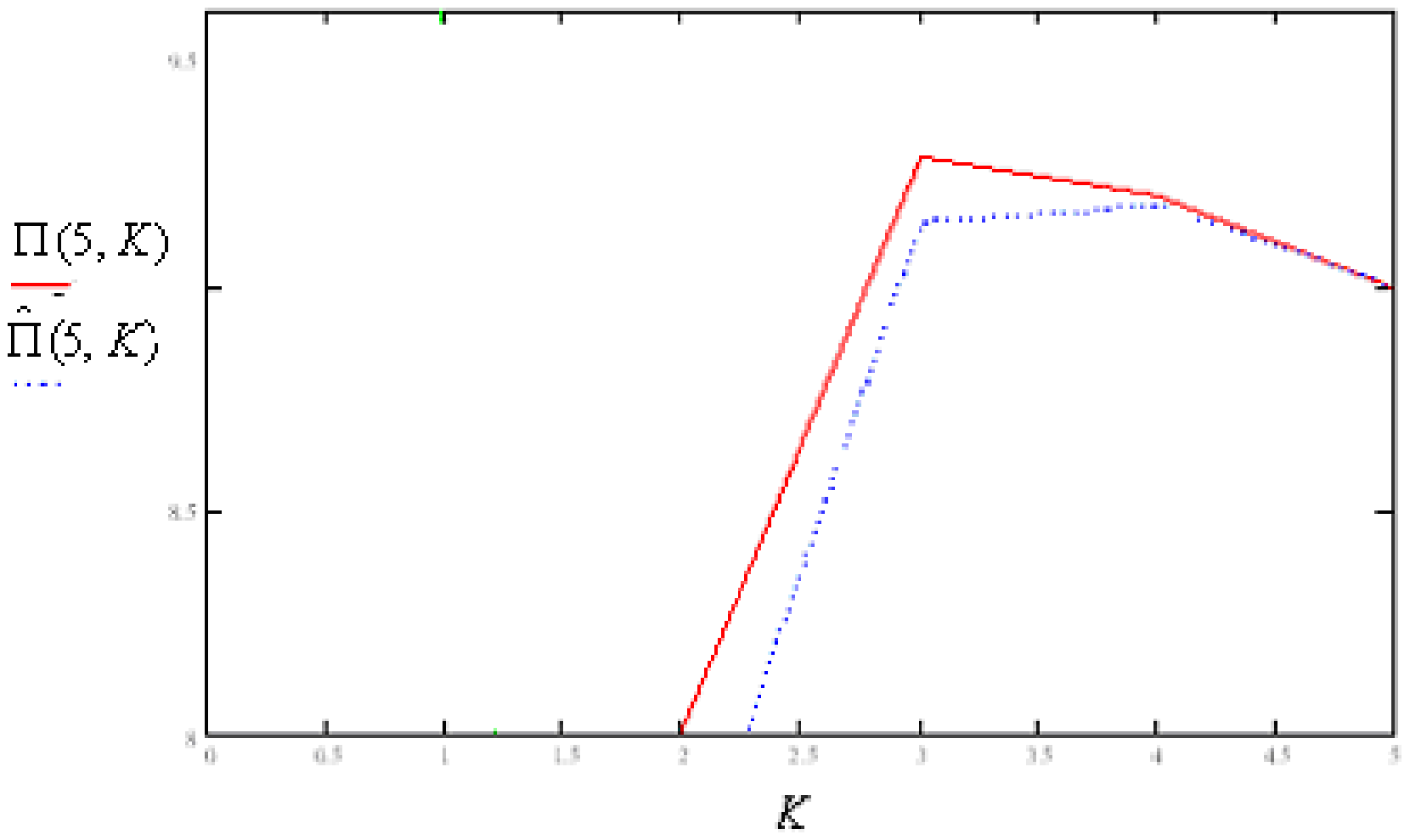

**Fig. 5 The optimal value of the average income, depending from initial inventory level *K***

*Inferences*

Resampling-approach can be successfully used for obtaining the estimators of parameters of interest of the renewal processes. Obtained formulas allow calculating the variance of the estimators for resampling and classical approaches. Numerical examples show the efficiency of suggested approach, taking estimators' mean squared error as efficiency criterion. This approach can be good alternative to traditional one and it can be applied for the inventory control of trusted, high reliable and expensive products.

## Conclusions

1. The work is devoted to the problems perfection of the organization and control of transport and logistic processes on base of intensive computer methods of statistics. The actuality of it is caused by the fact, that the adequate description of any process is necessary condition of it effective organization and control.
2. Resampling, which is the main intensive computer method of statistics, comparing with traditional methods has the following advantages: it is nonparametric it means that it is free from the mistakes connected with the choosing the incorrect distribution type of random variables; it is based on the direct imitation of the considered process, which leads to this method implementation in the situations, when the analytical description and analysis of the process are impossible.
3. The practical implementation of resampling approach distinguish with the singular simplicity, so the main purpose of the work consist in the investigation of its efficiency for the problems typical for transport and logistics. It is the theoretical part of the work. The efficiency of resampling approach were analyzed comparing with the traditional approach in the following tasks of mathematical statistics:
   - the robustness of the estimators of the multiple linear regression parameters;
   - the estimation of systems efficiency performance of the infinite servers queuing, which is widely implemented in reliability and insurance;
   - the comparison of two different renewal processes.

The analytical investigations and experimental results allow to conclude, that resampling approach is more preferable in the case of small initial sample sizes.

4. Appropriateness of the resampling approach usage is illustrated with the following practical tasks of the logistics and transport:
   - forecasting of the demand for aviation transport in the European Union countries, as the function of such factors: territory, population, gross domestic product per head, average month salary of one inhabitant.
   - forecasting of the reliability of the production with accumulation and development of damages;

- determining of the optimal initial level of the inventory in case of random demand and supply processes.

*5.* The obtained theoretical results are new and didn't found before in the literature. The methodic and the results of the forecasting of the demand on the passengers transportation by air were requested to be included in the report of the Second project of the Education and Science ministry: Mathematical methods development and estimation for the forecasting of passengers and freights flows in Baltic region.( *2. Izglītības un zinātnes ministrijas projektu: Matemātisko metožu izstrādāšana un novērtēšana pasažieru un krāvu plūsmu prognozēšanai Baltijas reģionā).*

6. The results of the work were presented at 9 International scientific conferences and seminars and were published in 10 articles. The author was the only author in 4 of them.